\documentclass[prb,aps,multicol,epsf]{revtex4}
\usepackage{graphicx}
\usepackage{epsfig}
\newcommand{\ba}{\begin{eqnarray}} 
\newcommand{\ea}{\end{eqnarray}} 
\newcommand{\be}{\begin{equation}} 
\newcommand{\ee}{\end{equation}} 
\newcommand{\bea}{\begin{eqnarray}} 
\newcommand{\eea}{\end{eqnarray}} 
\def\fat#1{{\bf #1}}
\def\etal{{\it et al}.}
\def\bra#1{{\left\langle#1\right\vert}}
\def\ket#1{{\left\vert#1\right\rangle}}

\def\jchi#1#2{{\chi}^{(#1)}_{(#2)}}
\def\jc#1#2{{c}^{(#1)}_{(#2)}}

\def\jl#1#2{{\rm \bf l}^{#1}_{#2}}

\def\kp{k^{\prime}}
\def\kpp{k^{\prime\prime}}
\def\cross{\times}
\def\refeqn#1{Eq.(\ref{#1})}
\def\refsect#1{Section \ref{#1}}

\def\sectref#1{Section \ref{#1}}

\def\dimtwo#1#2{#1 \AA~$\times$ #2 \AA}
\def\bigrod{\dimtwo{32.3}{45.5}}
\def\smallrod{\dimtwo{21.4}{24.79}}

\begin{document}

\def\CC{{\rm\kern.24em \vrule width.04em height1.46ex depth-.07ex \kern-.30em C}
}

\title{Tight-binding $g$ Factor Calculations of CdSe Nanostructures}

\author{Joshua Schrier and K. Birgitta Whaley}

\affiliation{Department of Chemistry and Pitzer Center for Theoretical
Chemistry, University of California, Berkeley 94720}

\begin{abstract}
The Lande $g$ factors for CdSe quantum dots and rods are investigated
within the framework of the semiempirical tight-binding method.  We 
describe methods for treating both the $n$-doped and neutral
nanostructures, and then apply these to a selection of nanocrystals of
variable size and shape, focusing on approximately spherical dots and
rods of differing aspect ratio.  For the negatively charged $n$-doped
systems, we observe that the $g$ factors for near-spherical CdSe dots
are approximately independent of size, but show strong shape
dependence as one axis of the quantum dot is extended to form rod-like
structures.  In particular, there is a discontinuity in the magnitude
of $g$ factor and a transition from anisotropic to isotropic
$g$ factor tensor at aspect ratio $\sim 1.3$.  For the neutral
systems, we analyze the electron $g$ factor of both the conduction and
valence band electrons.  We find that the behavior of the electron
$g$ factor in the neutral nanocrystals is generally similar to that in
the $n$-doped case, showing the same strong shape dependence and
discontinuity in magnitude and anisotropy.  In smaller systems the
$g$ factor value is dependent on the details of the surface model.
Comparison with recent measurements of $g$ factors for CdSe
nanocrystals suggests that the shape dependent transition may be
responsible for the observations of anomalous numbers of $g$ factors
at certain nanocrystal sizes.
\end{abstract}

\maketitle
 
\section{Introduction} \label{intro}

The electronic structure and linear optical spectroscopy of
semiconductor nanocrystals have been the subject of considerable
theoretical attention over the last ten years.  The size scaling of
excitonic absorptions, excitonic fine structure, and role of atomistic
effects such as surface reconstruction, are relatively well
understood.  Less is known about the behavior of the electronic states
in the presence of a magnetic field.  Recent experimental
demonstrations of long-lived spin coherences in semiconductor
nanostructures\cite{AK99} have provided motivation for a detailed
fundamental investigation of the behavior of electronic excitations in
magnetic fields. The spin lifetimes appear to be longest in
nanostructures possessing full three-dimensional confinement, namely
quantum dots, where undoped nanocrystals show room temperature spin
lifetimes of up to 3 ns,\cite{GAP+99} considerably larger than the
corresponding lifetimes for undoped quantum wells and bulk
semiconductors ($\sim 50$ psec - 1 nsec).\cite{AK99} Since lifetimes
are typically significantly increased by doping, quantum dots show
considerable potential for optimizing long-lived spin degrees of
freedom.\cite{WAB+01}

Despite this experimental promise, to date even the basic
magneto-optical phenomena in these nanostructures are not well
understood theoretically.  Thus, one unexplained phenomenon in the
study of semiconductor nanocrystals (NC's) is the appearance of
multiple Lande $g$ factors in CdSe quantum dots (QD's).  Time Resolved
Faraday Rotation (TRFR) studies of excitons in CdSe QD's reveal
multiple $g$ factors for certain dot sizes, with either two or four
values detected.\cite{GAP+99,GAE+02,Gup02} However, magnetic circular
dichroism (MCD) measurements apparently reveal a single $g$ factor per
exciton state in the two dot sizes studied (19 \AA~ and 25 \AA~
diameter).\cite{KNB+98} As noted in Ref.~\onlinecite{GAP+99}, the
Faraday rotation in neutral quantum dots should contain signatures of
both electron and hole spins, with the relative contributions
determined by the detailed coupling between these in the excitonic
state.  However there is currently no real understanding of why more
than one $g$ factor should be observed in TRFR, nor why for one
particular size four values are detectable.  Some investigators have
conjectured that multiple $g$ factors may result from excitonic fine
structure in the QD energy levels, reflecting the fact that different
fine structure levels would be expected to possess different excitonic
$g$ factors.\cite{KNB+98} The bright and dark exciton states for CdSe
have been predicted to possess quite different excitonic $g$ factor
values.  However fits to magnetic field-dependent
polarization-resolved photoluminescence spectra to extract the
$g$ values for the dark exciton states\cite{J-HA+01} do not show
agreement with corresponding measurements from circular
dichroism.\cite{KNB+98} Moreover, TRFR measurements with excitation
energies tuned to different excitonic fine structure states do not
appear to show different $g$ factors.\cite{GAP+99} Others have
proposed that both the electron and exciton signatures may all be
present in TRFR,\cite{GAP+99} or that multiple values arise from an
electronic contribution coexisting with an exciton contribution within
an ensemble of QDs.\cite{GAE+02} Comparison of values extracted from
MCD and from TRFR is not straightforward; whereas the effective mass
treatment of MCD experiments calculates the exciton $g$ factor from a
constant electron contribution and a calculated hole
$g$ factor,\cite{KNB+98} the treatment of what is hypothesized to be
an exciton $g$ factor in TRFR experiments is obtained using a
calculated electron contribution and uses a fixed hole
$g$ factor.\cite{GAE+02}

Theoretical analysis is complicated by the possible contributions of
crystal symmetry induced anisotropy in the hole $g$ factor, by the
effects of exchange coupling in exciton states, as well as by the
possible effects of nanocrystal shape and surface contributions.
Conversely, experimental efforts to assign the multiple $g$ factors
observed are complicated by the use of a distribution of randomly
oriented nanocrystals having non-uniform size and shape.  Efforts to
average anisotropy that might arise in a TRFR-measured exciton
$g$ factor as a result of anisotropy of the hole $g$ factor in a
hexagonal crystal over an ensemble of randomly oriented QDs do not
show qualitative agreement with experiment.\cite{Gup02} A recent
proposal based on effective mass analysis has suggested that exciton
precession is exhibited only by a special subset of the QD ensemble,
termed `quasi-spherical', in which there is an effective cancellation
between the intrinsic anisotropy due to the hexagonal structure and
that due to the nanocrystal shape, and resulting in isotropic
$g$ factors, while all other shape QDs presumed to exhibit only electron
precession.\cite{GAE+02} We shall term this the isotropically
quasi-spherical region, since no explicit aspect ratio range is
proposed and the use of the term `spherical' refers only to the
three-fold degeneracy of the $g$ tensor components.  This regime is to
be distinguished from the geometrically quasi-spherical region in
which the QDs have aspect ratios near unity.  In general, the
isotropically and geometrically quasi-spherical regimes may or may not
be coincident.

In this work we investigate the $g$ factors of CdSe nanostructures
within the framework of semiempirical tight-binding.  Unlike effective
mass treatments,\cite{KKI99} this has the advantage of retaining the
atomistic nature of the problem, and thereby allowing for realistic
treatment of ligand and reconstruction effects at the
surface.\cite{PW99} Since it is possible to synthesize a wide variety
of CdSe nanostructures, such as rods and tetrapods, in a controlled
fashion,\cite{MSA00} it is useful to have a theory which may be
applied to arbitrary structures.  To demonstrate this point, we
describe here calculations for both CdSe dots and for rods of variable
aspect ratio.  Additionally, it is possible to create electrically
$n$-doped dots,\cite{SG-S00,WSG-S01} although the $g$ factors for
these systems have not yet been experimentally determined.  To a first
approximation, the $n$-doped electron $g$ factors will be equivalent
to those of an electron in the conduction band.  In light of this, we
describe here theoretical treatments for both $n$-doped and excitonic
systems.  We have endeavored to use only tight-binding parameters
applied previously in the literature to treat other properties.  In
particular, we employ a tight-binding description that was augmented
for linear optical properties.\cite{LPW98}  We note that while this
use of a parameterization that has not been optimized specifically for
magneto-optical properties may limit our ability to obtain
quantitatively accurate results, the qualitative physical behavior
should nevertheless give us an understanding of the effects of NC size
and shape on the anisotropy of the $g$ tensor.  The calculations in
this paper employ tight-binding models of nanocrystals possessing
unreconstructed surfaces, using realistic models of surface
passivation developed previously.\cite{PW99,LPW98,LW99} The
modifications that might be induced by surface reconstruction will be
briefly discussed in the context of calibration calculations made with
truncated surfaces.

\section{Theory}
\subsection{Single particle Hamiltonian} \label{hamiltonians}

The effective single-particle Hamiltonian is calculated with the nearest-
neighbor
$sp^{3}s^{*}$ basis tight-binding approach, using the standard semiempirical
matrix elements,\cite{LL90} transformed from zinc blende to 
hexagonal crystal structure according to the transformations 
given in Ref.~\onlinecite{PW99}.  
Numerical diagonalization yields
the single particle states, $\psi_{i}({\fat{r}}_{i})$, and
single particle energies, $E_{i}$ for a given state $i$.  

\subsection{$g$ tensor for a single electron in the conduction level}\label{ndoped}

The derivation of the $g$ factor in finite molecular systems is conducted by
equating the phenomenological spin Hamiltonian,
\be
H_{phenom} = \mu_{B} {\bf B} \cdot {\bf g} \cdot {\bf s},
\ee
(where $\mu_{B}$ is the Bohr magneton, ${\bf B}$ is the magnetic field
vector, $ {\bf g}$ is the so-called $g$ tensor, and ${\bf s}$ is the
spin-vector), with the theoretical spin Hamiltonian
\be
H_{spin} = g_{0}{\mu}_{B}{\bf s}\cdot{\bf B} + \xi {\bf l}\cdot{\bf s}
+ {\mu}_{B}{\bf l}\cdot{\bf B}.
\ee
Here $g_{0}$ is the free-electron $g$ factor, $\xi$ is the
spin-orbit coupling, and ${\bf l}$ is the orbital-angular momentum
operator).  We have neglected the contribution of hyperfine interactions here.  
This spin Hamiltonian $H_{spin}$ containing
all magnetic field and spin-orbit coupling terms, is to be added to a spatial 
Hamiltonian, $H_0$, that is evaluated here within tight-binding.\cite{LL90}

Our method for calculating the $g$ tensors of $n$-doped systems
follows the theoretical treatment made for finite molecular systems 
by Stone, which treats $H_{spin}$ as a second order perturbation.\cite{Sto63} 
Similar
 Extended H\"uckel treatments of
organometallic compounds\cite{KdVvdA72} and small
radicals\cite{Min74} have also been reported.  The
$g$ tensor for a doublet radical, corresponding to a single unpaired electron 
spin, is given by 
\be\label{g_doublet}
g_{ij} =
g_{0}\delta_{ij}+2\sum_{k,n\ne0}{{\bra{\psi_{0}}\xi_{k}(r_{k})\jl{i}{k}\ket{\psi
_{n}}{\bra{\psi_{n}}}\jl{j}{k}\ket{\psi_{0}}}
\over {E_{0} - E_{n}} }, 
\ee
where $\{i,j\}$ are Cartesian components, $g_{0}$ is the free
electron $g$ factor, $\psi_{0}$ denotes the single-particle 
eigenvector corresponding to the
unpaired electron state, $n$ runs over all of the doubly-occupied
and virtual orbitals, $\xi_{k}(r_{k})$ is the spin-orbit
coupling as a function of $r_{k}$, and $\jl{i}{k}$ is the orbital angular 
momentum operator
component in the $i$th Cartesian direction centered on the $k$th atom.
We assume that 
the additional doping electron
can simply be placed in the lowest unoccupied molecular orbital that is derived 
from
the single-particle calculation.
 We have also neglected the gauge-correction term,\cite{Sto63} 
e.g., for $g_{zz}$,
\be
{{m}\over{\hbar^{2}}} \bra{\psi_{0}}\sum_{k}\left (
{x}^{2}_{k}+y_{k}^{2} \right ) \xi_{k}(r_{k}) \ket{\psi_{0}}.
\ee
This is justified since both Stone's analysis and our own preliminary 
calculations
indicate that the magnitude of this term is small.  However, as
discussed at the end of this section, as well as in
\sectref{validity}, our estimation of this term is dependent
on the magnitude of the transition dipole matrix
elements, and hence dependent on the parameterization
of these values.

We expand the $n$-th single-particle state in terms of the basis of atomic orbitals
at the $k$ site,
\be\label{expandnewnotation} 
\ket{\psi_{n}}=
\sum_{k}\ket{\jchi{n}{k}} = \sum_{k} \left \lbrace \jc{n}{s,k}\ket{s,k} 
+ \jc{n}{p_{x},k}\ket{p_{x},k} + \jc{n}{p_{y},k}\ket{p_{y},k}
+ \jc{n}{p_{z},k}\ket{p_{z},k} + \jc{n}{s^{*},k}\ket{s^{*},k} \right \rbrace.
\ee
Here the ket form of $\jchi{n}{k}$ is to be understood as consisting of all the
atomic ($\lbrace sp^{3}s^{*}\rbrace$)
orbitals on site $k$, with coefficients included according to the expression in
Eq.~(\ref{expandnewnotation}).  We now
introduce two approximations to simplify the evaluation of the
spin-orbit 
coupling and  orbital angular momentum matrix elements in 
\refeqn{g_doublet}, following Stone.\cite{Sto63}  First, since the
spin-orbit coupling is $\xi(r) \sim r^{-3}$, then $\xi_{k}(r_{k})$ is
effectively zero except near atom $k$, and thus
\be\label{ls_coupling}
\bra{\psi_{0}}\xi_{k}(r_{k})\jl{i}{k}\ket{\psi_{n}} \approx 
\bra{\jchi{0}{k}}\xi_{k}(r_{k})\jl{i}{k}\ket{\jchi{n}{k}} \approx
\xi_{k}\bra{\jchi{0}{k}}\jl{i}{k}\ket{\jchi{n}{k}}, 
\ee
where $\xi_{k}$ is the spin-orbit coupling constant for atom $k$, which has
been parameterized for semiconductor systems by Chadi:\cite{Cha77} 
$\xi_{Cd}=0.151\,{\rm eV}$, $\xi_{Se}=0.32\,{\rm eV}$.
Since, in our current model for CdSe, the oxygen ligands are
modeled as consisting of an $s$-orbital only,\cite{LPW98} they contribute no
spin-orbit coupling.  This may be justified by noting that $\xi_{O}=0.0187\,{\rm eV}$
is an order of magnitude smaller than $\xi_{Cd}$.\cite{MRW62}
Second, for the orbital angular momentum matrix element,
\be\label{stonesfirstapprox}
\bra{\psi_{n}}\jl{j}{k}\ket{\psi_{0}} =
\sum_{\kp,\kpp}
\bra{\jchi{n}{\kp}}\jl{j}{k}\ket{\jchi{0}{\kpp}},
\ee
using the relation $l_{k} = l_{\kp} + {\hbar}^{-1}{{\bf r}_{k
\kp}}\cross\fat{p}$.  Assuming that the atomic orbitals are
approximate eigenfunctions of parity, we can show that for 
\refeqn{ls_coupling} not
to vanish, the matrix elements of $\fat{p}$ must vanish.  In addition,
a tight-binding treatment assumes that overlap between orbitals on different 
atoms
is zero, leading to
\be\label{l_approx}
\bra{\psi_{n}}\jl{j}{k}\ket{\psi_{0}} \approx 
\sum_{\kp}\bra{\jchi{n}{\kp}}\jl{j}{\kp}\ket{\jchi{0}{\kp}}.
\ee
Combining these two approximations, \refeqn{g_doublet} becomes
\be \label{g_doublet_approx}
g_{ij} =
g_{0}\delta_{ij}+2\sum_{n\ne0}
{{ { \left ( \sum_{k} \xi_{k} \bra{\jchi{0}{k}}\jl{i}{k}\ket{\jchi{n}{k}}
\right )}
{\left ( \sum_{\kp} \bra{\jchi{n}{\kp}}\jl{j}{\kp}\ket{\jchi{0}{\kp}}
\right ) } } \over {E_{0} - E_{n}}} . 
\ee
We have used $\vert E_{0} - E_{n} \vert < 0.05 \,{\rm meV}$ as
the criterion for degeneracy in the calculations reported here.

Though the first approximation, 
Eqs. (\ref{ls_coupling}) - (\ref{stonesfirstapprox}), 
is quite reasonable, one might question
the validity of the second approximation, \refeqn{l_approx}.  Indeed, 
in semiempirical Intermediate Neglect of Differential Overlap (INDO) type
calculations, 
\cite{TUK+97,HZ99,BPM+00} it has
been found numerically that this is not a good approximation.  For the
general element $\bra{\jchi{}{k}}\jl{i}{\kp}\ket{\jchi{}{\kpp}}$, there
are five different equality or
nonequality relations between $k$, $\kp$, and $\kpp$ (i.e., $k = \kp =
\kpp$, $k = \kp \neq \kpp$, $k = \kpp \neq \kp$, $\kp = \kpp \neq k$,
$k \neq \kp \neq \kpp$).  
Evaluating these in the atomic basis, 
using the relations $l_{k} = l_{\kp} + {\hbar}^{-1}{{\bf r}_{k\kp}}\cross\fat{p}$ 
and $p_{\alpha} = i m_{e} {\hbar}^{-1}\left
[H,\alpha\right ]$, and taking the overlap between orbitals on different atoms 
as
zero (which is an assumption of the tight-binding formalism
itself and not introduced in our derivation), then the correction term that must 
be added
to the $\alpha$ component ($\alpha,\beta,\gamma$ Cartesian components)
of \refeqn{l_approx} is
\be\label{stone_correction}
\sum_{\kp\in nn(k)} \sum_{\kpp \in nn(k) \atop \kpp \ne \kp} 
\epsilon_{\alpha\beta\gamma}
\left ( {{m_{e}} \over {i \hbar^{2}}} \right ) r^{(\beta)}_{k\kpp} 
\bra{\jchi{n}{\kp}}\left [r^{(\gamma)}, H_0\right ] \ket{\jchi{0}{k}},
\ee
where $ \epsilon_{\alpha\beta\gamma}$ is the Levi-Civita symbol, 
$nn(k)$ indicates nearest neighbors of the $k$-th atom,
$m_{e}$ is the electron rest mass, $r^{(\beta)}_{k\kpp}$ is the
$\beta$ component of the distance between atoms $k$ and $\kpp$,
$r^{(\gamma)}$ is the $\gamma$ component of the position operator, and
$H_0$ is the tight-binding Hamiltonian.  In the present tight-binding 
description that is 
augmented for optical properties, the matrix elements of $r^{(\gamma)}$ are
obtained from the empirical transition dipole matrix
elements.\cite{LPW98} We have calculated the effect of this correction on
the $g$ factor for all dots and rods.  In no case did it make a
difference of more than $10^{-5}$ in the $g$ factor.  Since this is
beyond the inherent accuracy of tight-binding, we conclude that this
contribution may be omitted, substantially reducing the computational cost.

\subsection{Electron $g$ tensor for a pair of electrons in separate levels}\label{excitonicgtheory}

In the case of neutral, undoped nanocrystals, unpaired spin (leading
to a measurable electron spin resonance (ESR) signal) is caused by the
creation of an exciton.  It is often stated that in a particular NC, one
may observe either an electron $g$ factor due to only the conduction
band electron as a result of rapid hole dephasing, or else a single $g$ factor 
that results from exciton
precession.\cite{RER+02,GAE+02}  
The hole is assumed not to precess and its $g$ factor is used as a fitting parameter in
the effective mass expression for the exciton $g$ factor.\cite{GAE+02}  
The electron $g$ factor in this situation
is to a first approximation, the same as the conduction electron
$g$ factor calculated in the previous section.

However, there is also the possibility, analogous to molecular ESR, in which
an electron $g$ factor is observed which is due to multiple unpaired
electrons.  For an arbitrary number of unpaired spins in single-electron levels
$p$,
having total spin $S$, \refeqn{g_doublet} becomes \cite{Sto63}
\be\label{garbitrary}
g_{ij} =
g_{0}\delta_{ij}+{{1}\over{S}}\sum_{p}\sum_{n\ne p}\sum_{k}
{{\bra{\psi_{p}}\xi_{k}(r_{k})\jl{i}{k}\ket{\psi_{n}}{\bra{\psi_{n}}}\jl{j}{k}
\ket{\psi_{p}}}
\over {E_{p} - E_{n}} }.
\ee
Only the electron configuration of highest multiplicity is observed in ESR
(and thus treated here), since the others have non-zero electric 
dipole matrix elements with lower multiplicity configurations.  In our case, this
corresponds to the $S=1$ triplet (``dark'') exciton-like state, with
neglect of electron-hole correlation.

We remark that this approach is not generally applicable to the
treatment of the exciton $g$ factor.  In general, one needs to consider
the effect of the magnetic field on the total angular momentum $J=L+S$
of the particle.  However, in the $n$-doped case our state is a
nondegenerate doublet ground state and as such may be represented by a
real wavefunction; as a result $\langle L \rangle=0$, allowing us to perform the
treatment above.  
This
does not hold if one considers mixing with excited electronic
configurations.  Taking into account the effect of magnetic field on $J$
with a non-perturbative treatment of $H_{spin}$ leads to similar trends in
the anisotropy, however with significantly lower magnitude of electron
g-factors, in better quantitative agreement with experiment. 
[P. C. Chen and K. B. Whaley, to be published.]

\section{Results}
\subsection{Building Nanocrystals}

The crystal structures used here for the dots are the same as those used
in previous tight-binding studies.\cite{LPW98,PW99}  These crystals are facetted 
with $C_{3v}$ symmetry.  They
incorporate ligand effects through a semiempirical oxygen-like
``atom'' that fully saturates the cadmium surface sites.  The surface selenium 
atoms possess dangling bonds.  A set of calibration calculations that removed 
the dangling bonds, {\it i.e.}, truncated the surface Se atoms, was
also performed in order to 
provide some assessment of the effect of dangling bonds on the magnitude of the 
$g$ factors.    To construct the nanocrystal rods, we used 
the largest dot as a template for the crystal structure, and then removed 
successive layers of the sides parallel to 
the non-wurtzite axes in order to arrive at the desired
rod diameter.  Surface ligands were added using the hydrogen-addition function
in PC Spartan 2002, and then the Cd-ligand bond lengths were
lengthened 
to 2.625\, \AA~ using a
perl script. Two 
series of rods were studied, possessing smaller or larger cross-sections.  The 
first (smaller) series has diameter
\smallrod, and the second (larger) series has diameter 
\bigrod. Note that since the crystal is in fact
hexagonal, two dimensions are required to completely specify the cross-section 
of each rod, although we shall loosely distinguish the two series by their 
effective ``diameters''.
In both series, the shorter rods were created by removing two planes (of total 
width 3.5\,\AA)
of atoms perpendicular to the wurtzite axis, in addition to the removal of 
layers from the sides parallel to this axis. This additional removal was 
necessary to
keep the surface characteristics similar on all rod surfaces.  
This procedure for creating rods 
removes the $C_{3v}$ symmetry of the nanocrystals, but does result in faceted 
rods possessing shapes that are qualitatively in agreement with the shapes 
characterized experimentally by transmission electron microscopy.\cite{CBK+97} 
Additionally, in 
order to lengthen
the rods, the structure of the preceeding 7.0 \AA~ in the wurtzite axis
direction was duplicated and translated.   
Figure 1 shows cross-sections for the two series of rods employed here.  
Cross-sections of the $C_{3v}$ symmetry dots are shown in Ref. ~\onlinecite{LPW98}.

\subsection{$n$-doped systems}
\subsubsection{Dots}\label{dopeddotresults}
We calculated the $g$ tensor for $n$-doped dots with diameters ranging
from approximately 17 \AA~- 50 \AA.  Note that since these 
calculations are atomistic, the dots are faceted and are therefore
only approximately spherical.   The effective sphericity is given in 
Ref.~\onlinecite{LPW98}. 
The $g$ factor results for oxygen-passivated nanocrystals are
shown in Figure 2a. We found the
$g$ factors to be relatively size-independent, and to possess average 
$g$ factor values of $\sim 2$. 
  We have 
determined values for the anisotropic components as $g_{\parallel} \simeq 2.010$ and
$g_{\perp} \simeq 2.004$.  These components are identified by their 
degeneracy, with $g_{\perp}$ being two-fold degenerate, and $g_{\parallel}$ 
being singly degenerate.  This small value of anisotropy would be extremely 
difficult to resolve experimentally, and the near-spherical shape dots are thus 
expected to appear ``isotropic''. 

\subsubsection{Rods} \label{dopedrodresults}
We calculated the $g$ tensor for $n$-doped rods of diameters \bigrod~
and \smallrod~ for various lengths, shown in Figures 2b and 2c-d,
respectively.  We found that in both cases the $g$ factor changes
abruptly when the length of the rod is approximately 1.3 times the
diameter.  The anisotropic components in both cases experienced a
similar discontinuity.  For the smaller diameter (\smallrod) rod, the
discontinuity is between $g_{iso} = 1.998 \pm 0.023$ (from 14.0 \AA~ -
28.0 \AA) and $g_{iso} = 1.913 \pm 0.020$ (from 31.5 \AA~ - 45.5 \AA),
or $\Delta g_{iso} = 0.085 \pm 0.043$ between the 28.0 \AA~ and 31.5
\AA~ crystals.  This region is shown as an inset in Figure 2c.  Note
that since this is a discrete atomistic treatment, we cannot ``cut''
the crystal at distances less than the 3.5 \AA~ spacing.  For the
larger diameter (\bigrod) rod, we calculated $\Delta g_{iso} = 0.3$,
with the discontinuity in the isotropic $g$ factor at approximately
1.3 times the smaller dimension of the diameter (between 38.5 \AA~ and
42.0 \AA).  Since this dot-rod transition discontinuity is at least an
order of magnitude greater than the TRFR
resolution,\cite{GAE+02,Gup02} it should be possible to measure this
effect and could provide a useful method of examining aspect ratios
during nano-rod synthesis. Furthermore, it is worth noting the large
deviation from the free electron $g$ factor and large $g$ anisotropy
that is possible by manipulation of shape alone, as demonstrated in
the highly elliptical dots (Figures 2b and 2c).  In the case of the
larger rods, we observe that the $g$ factor is anisotropic for the
dot-like structures, then becomes essentially isotropic for crystals
between the length of 42 \AA~ to 71 \AA, and then becomes anisotropic
again for longer rod structures.  This appears to bound the
isotropically quasispherical region between aspect ratios $1.3-2$.  In
the case of the smaller rods, both the isotropic and anisotropic
components experience large changes as a function of size.  This is to
be expected when making a size study of small structures based on an
atomistic model, since adding a layer of atoms to a small system
provides a large perturbation of shape.

\subsection{Neutral Systems}
\subsubsection{Dots}

We evaluated the $g$ factor for an electronic configuration with
parallel spin electrons in conduction and valence band edge states 
(which we will refer to as the ``excitonic electron'' $g$ factor) for
each of the dot sizes treated in \sectref{dopeddotresults} using the approach
discussed in \sectref{excitonicgtheory}.  Examining the
results in Figure 3a, the largest of the dots treated theoretically here is
comparable to the smallest dot treated
experimentally, but there is no quantitative
agreement with the experimentally measured $g$ factors of $1.63 \pm 0.01$ 
and $1.565 \pm 0.002, 1.83 \pm 0.01$ for the 40 \AA~ and  50 \AA~ diameter dots,
respectively.\cite{GAE+02,Gup02,GAP+99} However,
qualitatively, we
note that there are no anisotropies larger than a factor of
approximately $0.1$.

\subsubsection{Rods}

The excitonic electron $g$ factors for
the rods treated in \sectref{dopedrodresults} are shown in
Figures 3b and 3c.  The
discontinuity at aspect ratio 1.3 is still present but is 
reduced by an order of magnitude
for both rod sizes as compared to the $n$-doped electron $g$ factor.
However, the
presence of an isotropically quasi-spherical region is the same for
the electron in the excitonic system as for the $n$-doped electron $g$ factor.

\subsection{Truncated Surface Calculations}

Surface reconstruction is an important factor in the optical
spectroscopy of small NCs. \cite{LW99}  One result of surface
reconstruction for CdSe nanocrystals passivated by oxygen ligands is
to move the Se dangling bonds away from the band edge to lower
energies.  To a first approximation,
this can be modeled by removing the dangling 
selenium bonds on the
surface of the NC.  To ascertain the qualitative effect of surface
reconstruction on our results, we performed the calculations for
truncated nanocrystals without
the dangling selenium bonds.  Results are shown in Figures 4 and 5 for
the $n$-doped electron and excitonic electron respectively.  

Overall, we found the effect of surface truncation to be a decrease in
magnitude of the $g$ factor.  
For dots and for the \bigrod~ diameter rods, 
the behaviour of the $g$ factor components is qualitatively
similar for both dangling and non-dangling cases.
The apparent degeneracy of two of the $g$ components in the dangling
bond calculations (Figures 2b and 3b) is broken in the truncated
calculations  (Figures 4b and 5b).
For the smaller, 
\smallrod~ diameter rods, (Figures 4c and 5c)
the behavior of the $g$ components for surface truncated systems 
is qualitatively different.
In particular, the $g$ factor becomes isotropic at aspect ratio $\sim
3$ for both the $n$-doped (Figure 4c) and excitonic (Figure 5c) electron
$g$ factors.  Since one would expect the smaller crystals to show a more
profound change due to their larger surface area to volume ratio, it is
not entirely surprising that the behavior of small rods deviates from
that of larger rods.
The results suggest that surface reconstruction is an important
effect for the $g$ factors of small nanocrystals in both
quantitative and qualitative terms, and warrants more detailed investigation.

\subsection{Orbital Character}\label{orbitalchar}

To examine the origin of the discontinuity in the $g$ factor at aspect
ratio $1.3$, the
appearance of isotropic regions, and the general qualitative behavior
of $g$,
we examined the character of the near band-edge orbitals.  For the
conduction band edge state and nine states above as well as the 
valence band edge state and
nine states below, we calculated the fractional contribution of the
various types of atomic orbitals 
to the given molecular orbital.  
The results for the \bigrod~ rod are shown
in Figure 6.
The left panels show the orbital contributions
with the inclusion of dangling Se surface bonds, and the right panels
show the results from truncated nanocrystals with the dangling Se
bonds removed.  Shown within each
figure are graphs for the orbital contributions where the maximum
fractional content exceeded 0.15.  Dotted lines depict the 
the conduction band edge and higher states, and solid lines depict the
valence band edge and lower states.

Qualitatively, the fractional orbital
content of the conduction and
valence band edge states behave similarly for both types of surface treatments.
There is an increase in the Cd-$s$ contribution at aspect
ratio $\approx 1.3$, which then decreases at aspect ratio $2.5$, corresponding
to a simultaneous decrease of the Se-$p$ contributions.
While the behavior of the valence and conduction band edge states themselves are
relatively unaffected by the
surface treatment, truncating the dangling Se surface bonds appears to be a
reduction in the Se-$p$ content for the other states.
This is not surprising, in light of the similarity between the
$g$ factor behavior for the dangling and truncated calculations.

Results for the smaller rods are more complicated, and are not
shown.  Although the Cd-$s$ atomic orbital contribution is similar for both surface
treaments, the Se-$p$ level is qualitatively different.

\section{Discussion}

\subsection{Shape-Controlled $g$ factor Discontinuity}\label{discont_discuss}

\subsubsection{Relation to HOMO/LUMO Wavefunction}
The result concerning the discontinuity in the $g$ factor for CdSe
rods at the 1.3 aspect ratio suggests that small size differences
in the growth axis length  can have
large effects on both the magnitude and
the anisotropy of the $g$ factor.  An
examination of CdSe
rods using the semiempirical pseudopotential method by Hu
\etal\cite{HWL+02} studied changes in the electronic states as a
function of rod length.  In particular, level 
crossing occurs between the two highest occupied orbitals (i.e., the
HOMO and the level below it, HOMO-1) at an aspect ratio of $\sim 1.3$, and of
the HOMO-4 and HOMO-5 levels at an aspect ratio of $\sim 2$.  In each
case this level crossing involved a change in relative contributions of Se 
$4 p_{z}$ and Se $4 p_{x,y}$ levels, matching linearly polarized
emission spectroscopic results.\cite{HLY+01}  We observe qualitative agreement with
these results, as discussed in \sectref{orbitalchar}.  
However, since the pseudopotential study of Hu \etal , does not include surface 
reconstruction effects, the
details for small nanocrystals may differ.
It is obvious, however, that these changes in the orbital arrangement will
have a large effect on the $g$ factor, as it is dependent on the
orbital angular momentum of the state in question. 

Additionally, we have performed calculations in which we turned off
the wurtzite crystal field correction in our non-truncated surface
calculations 
to assess the role of the
crystal symmetry on the $g$ factor discontinuity.  For both rods, this
resulted in splitting the approximately degenerate $g$ levels in the
regions outside of the range of aspect ratio $1.3-2$, but the
existence of an isotropic region as well as the discontinuity in the
isotropic $g$ factor persisted.  This suggests that the discontinuity
and isotropic region that we observe are shape effects, rather than
simply a cancellation of the wurtzite crystal field, as proposed in
the ``quasi-spherical'' model.\cite{GAE+02}

\subsubsection{Connection to Experimental Observation of Multiple $g$ factors}

We conjecture that this discontinuity effect may play a role in the
existence of four $g$ factors in the experiments on 57-\AA~ radius
quantum dots.  This size dot is unique in showing four $g$ factors:
both slightly smaller and larger dots display only
two.\cite{GAP+99,GAE+02,Gup02}  It is well known that the so-called
``dots'' are in fact elliptical; empirically observed relations for
the ellipticity of quantum dots as a function of size, based on
transmission electron microscopy data, give an aspect ratio of $\sim
1.34$ for the 57-\AA~ dot, whereas other dots have either smaller or
larger aspect ratios.\cite{LPW98, Kad97, GAE+02} Since the size
control is on the order of $\pm 5 \% $, this suggests that unlike the
other samples studied, the size distribution of the 57-\AA~ dot may in
fact span the discontinuity we observe here.  We have tabulated the
dot size, number of $g$ factor components observed, and aspect ratios
in Table \ref{gcomponent_table}.  This suggests two possible
situations that may give rise to four $g$ components.  The first
scenario assigns the components as resulting from an exciton and an
isotropic electron component (as assigned in effective mass
studies\cite{GAE+02,Gup02}) deriving from the portion of the NC
ensemble in the isotropically quasispherical region, plus two
anisotropic electron components from the lower aspect ratio portion of
the ensemble.  The second possible assignment arises from one electron $g$ factor
and one exciton $g$ factor on either side of the discontinuity.  It is
our hope that this analysis will encourage TRFR experiments on even
more precisely size selected nanocrystal samples, as well as on
$n$-doped nanocrystalline systems, in order to distinguish between
these assignments.

\subsection{Extensions}\label{validity}

There are several limitations of this study.  The first is due to
 the use of a $sp^{3}s^{*}$ semiempirical basis.  In particular, the $s^{*}$
orbital was introduced by Vogl \etal\, with the intent of
mimicking $d$-orbitals.\cite{VHD83}  While satisfactory for optical
calculations, this orbital has no angular momentum, since $l = 0$ 
for $s^{*}$, as
opposed to $l = 2$ for $d$.  To go beyond this initial analysis, 
one might have to
include $d$-orbitals (i.e., use a $sp^{3}d^{5}$  or $sp^{3}d^{5}s^{*}$ tight-binding basis)
or else to include angular momentum for the $s^{*}$ orbital empirically.
Additionally, it is not clear that the semiempirical basis accurately
 corresponds to the eigenfunctions of angular-momentum that we
 attribute to it via $s$, $p$, etc., labels.  
Second, the ligand model treats oxygen as an $s$-orbital only,
neglecting any angular momentum contributions.  As mentioned in
\refsect{ndoped}, this is partially justifiable by the much smaller
spin-orbit coupling of oxygen compared to Cd or Se.  However, for small
crystals we expect this may fail, since the ratio of ligands to semiconductor
atoms increases.  
Again, it may be necessary to include a larger basis
(i.e., $p$-orbitals on the oxygen atoms)
or to determine an empirical correction to account for this effect.
Third, while we found the correction to Stone's second
approximation, \refeqn{stone_correction}, to be negligible, this is
dependent on the validity of the transition dipole matrix elements,
which were empirically devised to reproduce optical spectra,
\cite{LPW98}, and as a result may not be applicable to magneto-optical
problems.  
Fourth, the neglect of off-site terms in the evaluation
of the angular-momentum matrix elements
(Eqs. \ref{stonesfirstapprox}, \ref{l_approx}) further decreases the
magnitude of the shift from the free-electron $g$ factor.  If a more
quantitative analysis were desired, one could directly parameterize
these angular-momentum matrix elements by fitting to bulk or to
ab initio calculations of the $g$ factor in small clusters.  To our
knowledge, the latter has not been performed for CdSe, although there
exist separate studies of density functional theory (DFT) calculations on CdSe clusters of sizes
up to $\sim 200$ atoms\cite{TKC01} as well as methods to calculate the
$g$ tensor using DFT\cite{PZ01}. 
Finally, while treating the spin Hamiltonian perturbatively
is satisfactory in organic and organometallic molecules,[cite: 23]
this approximation may not be as appropriate for the quantitative
description of semiconductor systems, due to the larger spin-orbit
coupling constants.  Nevertheless, the qualitative trends with respect to
shape observed here do also appear to hold when the spin
 Hamiltonian is treated non-perturbatively. 
[P. C. Chen and K. B. Whaley, 
to be published.]

The issue of surface effects is complicated by the
shape dependence of the $g$ factor.  To proceed in future work, it may be most
effective to decouple these two effects.  To examine the effect of shape
alone (ignoring surface reconstruction effects), one may
 modify the existing effective mass treatments of the $g$ factor in
spherical nanocrystals to treat rods.  This would have the added
benefit of being able to treat the larger experimental nanocrystal
sizes, in particular the 57 \AA~ dot, to test whether the discontinuity
in the $g$ factor at aspect ratio 1.3 is present for larger size
crystals.  Second, since we have seen indications that surface reconstruction may have
substantial qualitative effects on the behavior of the $g$ factor in
small nanocrystals, one may apply the tight-binding
surface reconstruction method (via total energy minimization)
previously applied to CdSe nanocrystals,\cite{PW99} in order to resolve the
differences between the dangling Se-bond and truncated surface
calculations, and to determine whether this plays a role in why the smallest dot
studied in TRFR experiments shows only one $g$ factor 
component.\cite{GAE+02,Gup02}

\section{Summary}

We have developed a tight-binding theory for the Lande $g$ tensor for
electrons in $n$-doped and excitonic systems, which we have applied to
CdSe quantum dots and rods.  For $n$-doped systems, we found the
electron $g$ factor for approximately spherical dots to be independent
of dot size, while a discontinuity in the $g$ factor appears as the
c-axis is extended to form rod-like structures.  Similar behavior is
observed for excitonic electrons, although the magnitude of both the
$g$ factor and its discontinuity was found to be dependent on the
treatment of dangling surface Se bonds.  We also observe the existence
of a isotropically quasispherical regime between aspect ratio $1.3 -
2$ in all cases.  This appears to correspond to the ``quasi-spherical
hypothesis'' suggested in the effective mass treatments of the
$g$ factor.\cite{GAE+02} However, whereas the previous treatments
consider this as arising from the cancellation of wurtzite crystal
field effects on $g$ by shape terms, the isotropic region we observe
here appears to be due primarily to shape effects, and occurs even in
the absence of the wurtzite crystal field.  Comparison with available
experimental TRFR data indicates that the discontinuity between the
anisotropic and isotropic regions offers a possible explanation for
multiple $g$ factors.

{\it Note Added in Proof.} An
effective mass treatment for rod-shaped wurtzite nanocrystals has
recently been presented by Li and Xia, but
the method has not yet been applied to the calculation of $g$ factors.
[cite: X.-Z. Li and J.-B. Xia, Phys. Rev. B. {\bf 66}, 115316
(2002)]. 

\section{Acknowledgements}
We would like to thank Kenneth
Brown for many insightful conversations. J.S. thanks the National
Defense Science and Engineering Grant (NDSEG) program and U.S. Army
Research Office Contract/Grant No. FDDAAD19-01-1-0612 for financial
support.  K.B.W. thanks the Miller Institute for Basic Research in
Science for financial support.  This work was also supported by the
Defense Advanced Research Projects Agency (DARPA) and the Office of
Naval Research under Grant No. FDN00014-01-1-0826.

%\bibliography{cdsegfac}

\begin{thebibliography}{31}
\expandafter\ifx\csname natexlab\endcsname\relax\def\natexlab#1{#1}\fi
\expandafter\ifx\csname bibnamefont\endcsname\relax
  \def\bibnamefont#1{#1}\fi
\expandafter\ifx\csname bibfnamefont\endcsname\relax
  \def\bibfnamefont#1{#1}\fi
\expandafter\ifx\csname citenamefont\endcsname\relax
  \def\citenamefont#1{#1}\fi
\expandafter\ifx\csname url\endcsname\relax
  \def\url#1{\texttt{#1}}\fi
\expandafter\ifx\csname urlprefix\endcsname\relax\def\urlprefix{URL }\fi
\providecommand{\bibinfo}[2]{#2}
\providecommand{\eprint}[2][]{\url{#2}}

\bibitem[{\citenamefont{Awschalom and Kikkawa}(1999)}]{AK99}
\bibinfo{author}{\bibfnamefont{D.~D.} \bibnamefont{Awschalom}}
  \bibnamefont{and} \bibinfo{author}{\bibfnamefont{J.~M.}
  \bibnamefont{Kikkawa}}, \bibinfo{journal}{Physics Today}
  \textbf{\bibinfo{volume}{52}}, \bibinfo{pages}{33} (\bibinfo{year}{1999}).

\bibitem[{\citenamefont{Gupta et~al.}(1999)\citenamefont{Gupta, Awschalom,
  Peng, and Alivisatos}}]{GAP+99}
\bibinfo{author}{\bibfnamefont{J.~A.} \bibnamefont{Gupta}},
  \bibinfo{author}{\bibfnamefont{D.~D.} \bibnamefont{Awschalom}},
  \bibinfo{author}{\bibfnamefont{X.}~\bibnamefont{Peng}}, \bibnamefont{and}
  \bibinfo{author}{\bibfnamefont{A.~P.} \bibnamefont{Alivisatos}},
  \bibinfo{journal}{Phys. Rev. B} \textbf{\bibinfo{volume}{59}},
  \bibinfo{pages}{R10421} (\bibinfo{year}{1999}).

\bibitem[{\citenamefont{Wolf et~al.}(2001)\citenamefont{Wolf, Awschalom,
  Daughton, von Molnar, Roukes, Chtchelkanova, and Treger}}]{WAB+01}
\bibinfo{author}{\bibfnamefont{S.~A.} \bibnamefont{Wolf}},
  \bibinfo{author}{\bibfnamefont{D.~D.} \bibnamefont{Awschalom}},
  \bibinfo{author}{\bibfnamefont{R.~A.} \bibnamefont{Buhrmanand}},
  \bibinfo{author}{\bibfnamefont{J.~M.} \bibnamefont{Daughton}},
  \bibinfo{author}{\bibfnamefont{S.}~\bibnamefont{von Molnar}},
  \bibinfo{author}{\bibfnamefont{M.~L.} \bibnamefont{Roukes}},
  \bibinfo{author}{\bibfnamefont{A.~Y.} \bibnamefont{Chtchelkanova}},
  \bibnamefont{and} \bibinfo{author}{\bibfnamefont{D.~M.}
  \bibnamefont{Treger}}, \bibinfo{journal}{Science}
  \textbf{\bibinfo{volume}{294}}, \bibinfo{pages}{1488} (\bibinfo{year}{2001}).

\bibitem[{\citenamefont{Gupta et~al.}(2002)\citenamefont{Gupta, Awschalom,
  Efros, and Rodina}}]{GAE+02}
\bibinfo{author}{\bibfnamefont{J.~A.} \bibnamefont{Gupta}},
  \bibinfo{author}{\bibfnamefont{D.~D.} \bibnamefont{Awschalom}},
  \bibinfo{author}{\bibfnamefont{A.~L.} \bibnamefont{Efros}}, \bibnamefont{and}
  \bibinfo{author}{\bibfnamefont{A.~V.} \bibnamefont{Rodina}},
  \bibinfo{journal}{Phys. Rev. B} \textbf{\bibinfo{volume}{66}},
  \bibinfo{pages}{125307} (\bibinfo{year}{2002}).

\bibitem[{\citenamefont{Gupta}(2002)}]{Gup02}
\bibinfo{author}{\bibfnamefont{J.~A.} \bibnamefont{Gupta}}, Ph.D. thesis,
  \bibinfo{school}{University of California, Santa Barbara}
  (\bibinfo{year}{2002}).

\bibitem[{\citenamefont{Kuno et~al.}(1998)\citenamefont{Kuno, Nirmal, Bawendi,
  Efros, and Rosen}}]{KNB+98}
\bibinfo{author}{\bibfnamefont{M.}~\bibnamefont{Kuno}},
  \bibinfo{author}{\bibfnamefont{M.}~\bibnamefont{Nirmal}},
  \bibinfo{author}{\bibfnamefont{M.~G.} \bibnamefont{Bawendi}},
  \bibinfo{author}{\bibfnamefont{A.}~\bibnamefont{Efros}}, \bibnamefont{and}
  \bibinfo{author}{\bibfnamefont{M.}~\bibnamefont{Rosen}}, \bibinfo{journal}{J.
  Chem. Phys.} \textbf{\bibinfo{volume}{108}}, \bibinfo{pages}{4242}
  (\bibinfo{year}{1998}).

\bibitem[{\citenamefont{Johnston-Halpernin
  et~al.}(2001)\citenamefont{Johnston-Halpernin, Awschalom, Crooker, Efros,
  Rosen, Peng, and Alivisatos}}]{J-HA+01}
\bibinfo{author}{\bibfnamefont{E.}~\bibnamefont{Johnston-Halperin}},
  \bibinfo{author}{\bibfnamefont{D.~D.} \bibnamefont{Awschalom}},
  \bibinfo{author}{\bibfnamefont{S.~A.} \bibnamefont{Crooker}},
  \bibinfo{author}{\bibfnamefont{A.~L.} \bibnamefont{Efros}},
  \bibinfo{author}{\bibfnamefont{M.}~\bibnamefont{Rosen}},
  \bibinfo{author}{\bibfnamefont{X.}~\bibnamefont{Peng}}, \bibnamefont{and}
  \bibinfo{author}{\bibfnamefont{A.~P.} \bibnamefont{Alivisatos}},
  \bibinfo{journal}{Phys. Rev. B} \textbf{\bibinfo{volume}{63}},
  \bibinfo{pages}{205309} (\bibinfo{year}{2001}).

\bibitem[{\citenamefont{Kiselev et~al.}(1999)\citenamefont{Kiselev, Kim, and
  Ivchenko}}]{KKI99}
\bibinfo{author}{\bibfnamefont{A.~A.} \bibnamefont{Kiselev}},
  \bibinfo{author}{\bibfnamefont{K.~W.} \bibnamefont{Kim}}, \bibnamefont{and}
  \bibinfo{author}{\bibfnamefont{E.~L.} \bibnamefont{Ivchenko}},
  \bibinfo{journal}{Phys. Stat. Sol. B} \textbf{\bibinfo{volume}{215}},
  \bibinfo{pages}{235} (\bibinfo{year}{1999}).

\bibitem[{\citenamefont{Pokrant and Whaley}(1999)}]{PW99}
\bibinfo{author}{\bibfnamefont{S.}~\bibnamefont{Pokrant}} \bibnamefont{and}
  \bibinfo{author}{\bibfnamefont{K.~B.} \bibnamefont{Whaley}},
  \bibinfo{journal}{Eur. Phys. J. D} \textbf{\bibinfo{volume}{6}},
  \bibinfo{pages}{255} (\bibinfo{year}{1999}).

\bibitem[{\citenamefont{Manna et~al.}(2000)\citenamefont{Manna, Scher, and
  Alivisatos}}]{MSA00}
\bibinfo{author}{\bibfnamefont{L.}~\bibnamefont{Manna}},
  \bibinfo{author}{\bibfnamefont{E.~C.} \bibnamefont{Scher}}, \bibnamefont{and}
  \bibinfo{author}{\bibfnamefont{A.~P.} \bibnamefont{Alivisatos}},
  \bibinfo{journal}{J. Am. Chem. Soc.} \textbf{\bibinfo{volume}{122}},
  \bibinfo{pages}{12700} (\bibinfo{year}{2000}).

\bibitem[{\citenamefont{Shim and Guyot-Sionnest}(2000)}]{SG-S00}
\bibinfo{author}{\bibfnamefont{M.}~\bibnamefont{Shim}} \bibnamefont{and}
  \bibinfo{author}{\bibfnamefont{P.}~\bibnamefont{Guyot-Sionnest}},
  \bibinfo{journal}{Nature} \textbf{\bibinfo{volume}{407}},
  \bibinfo{pages}{981} (\bibinfo{year}{2000}).

\bibitem[{\citenamefont{Wang et~al.}(2001)\citenamefont{Wang, Shim, and
  Guyot-Sionnest}}]{WSG-S01}
\bibinfo{author}{\bibfnamefont{C.}~\bibnamefont{Wang}},
  \bibinfo{author}{\bibfnamefont{M.}~\bibnamefont{Shim}}, \bibnamefont{and}
  \bibinfo{author}{\bibfnamefont{P.}~\bibnamefont{Guyot-Sionnest}},
  \bibinfo{journal}{Science} \textbf{\bibinfo{volume}{291}},
  \bibinfo{pages}{2390} (\bibinfo{year}{2001}).

\bibitem[{\citenamefont{Leung et~al.}(1998)\citenamefont{Leung, Pokrant, and
  Whaley}}]{LPW98}
\bibinfo{author}{\bibfnamefont{K.}~\bibnamefont{Leung}},
  \bibinfo{author}{\bibfnamefont{S.}~\bibnamefont{Pokrant}}, \bibnamefont{and}
  \bibinfo{author}{\bibfnamefont{K.~B.} \bibnamefont{Whaley}},
  \bibinfo{journal}{Phys. Rev. B} \textbf{\bibinfo{volume}{57}},
  \bibinfo{pages}{12291} (\bibinfo{year}{1998}).

\bibitem[{\citenamefont{Leung and Whaley}(1999)}]{LW99}
\bibinfo{author}{\bibfnamefont{K.}~\bibnamefont{Leung}} \bibnamefont{and}
  \bibinfo{author}{\bibfnamefont{K.~B.} \bibnamefont{Whaley}},
  \bibinfo{journal}{J. Chem. Phys.} \textbf{\bibinfo{volume}{110}},
  \bibinfo{pages}{11012} (\bibinfo{year}{1999}).

\bibitem[{\citenamefont{Lippens and Lannoo}(1990)}]{LL90}
\bibinfo{author}{\bibfnamefont{P.~E.} \bibnamefont{Lippens}} \bibnamefont{and}
  \bibinfo{author}{\bibfnamefont{M.}~\bibnamefont{Lannoo}},
  \bibinfo{journal}{Phys. Rev. B} \textbf{\bibinfo{volume}{41}},
  \bibinfo{pages}{6079} (\bibinfo{year}{1990}).

\bibitem[{\citenamefont{Stone}(1963)}]{Sto63}
\bibinfo{author}{\bibfnamefont{A.~J.} \bibnamefont{Stone}},
  \bibinfo{journal}{Proc. Roy. Soc. A} \textbf{\bibinfo{volume}{271}},
  \bibinfo{pages}{424} (\bibinfo{year}{1963}).

\bibitem[{\citenamefont{Keijzers et~al.}(1972)\citenamefont{Keijzers, de~Vries,
  and van~der Avoird}}]{KdVvdA72}
\bibinfo{author}{\bibfnamefont{C.~P.} \bibnamefont{Keijzers}},
  \bibinfo{author}{\bibfnamefont{H.~J.~M.} \bibnamefont{de~Vries}},
  \bibnamefont{and} \bibinfo{author}{\bibfnamefont{A.}~\bibnamefont{van~der
  Avoird}}, \bibinfo{journal}{Inorg. Chem.} \textbf{\bibinfo{volume}{11}},
  \bibinfo{pages}{1338} (\bibinfo{year}{1972}).

\bibitem[{\citenamefont{Minaev}(1974)}]{Min74}
\bibinfo{author}{\bibfnamefont{B.~F.} \bibnamefont{Minaev}},
  \bibinfo{journal}{Opt. Spektrosk.} \textbf{\bibinfo{volume}{36}},
  \bibinfo{pages}{275} (\bibinfo{year}{1974}).

\bibitem[{\citenamefont{Chadi}(1977)}]{Cha77}
\bibinfo{author}{\bibfnamefont{D.~J.} \bibnamefont{Chadi}},
  \bibinfo{journal}{Phys. Rev. B} \textbf{\bibinfo{volume}{16}},
  \bibinfo{pages}{790} (\bibinfo{year}{1977}).

\bibitem[{\citenamefont{Morton et~al.}(1962)\citenamefont{Morton, Rowlands, and
  Whiffen}}]{MRW62}
\bibinfo{author}{\bibfnamefont{J.~R.} \bibnamefont{Morton}},
  \bibinfo{author}{\bibfnamefont{J.~R.} \bibnamefont{Rowlands}},
  \bibnamefont{and} \bibinfo{author}{\bibfnamefont{D.~H.}
  \bibnamefont{Whiffen}}, \emph{\bibinfo{title}{Atomic properties for
  interpreting {E}{S}{R} data}} (\bibinfo{year}{1962}),
  \bibinfo{note}{{National} Physical Laboratory \--- BPR 13}.

\bibitem[{\citenamefont{T{\"o}rring et~al.}(1997)\citenamefont{T{\"o}rring, Un,
  Kn{\"u}pling, Plato, and M{\"o}bius}}]{TUK+97}
\bibinfo{author}{\bibfnamefont{J.~T.} \bibnamefont{T{\"o}rring}},
  \bibinfo{author}{\bibfnamefont{S.}~\bibnamefont{Un}},
  \bibinfo{author}{\bibfnamefont{M.}~\bibnamefont{Kn{\"u}pling}},
  \bibinfo{author}{\bibfnamefont{M.}~\bibnamefont{Plato}}, \bibnamefont{and}
  \bibinfo{author}{\bibfnamefont{K.}~\bibnamefont{M{\"o}bius}},
  \bibinfo{journal}{J. Chem. Phys.} \textbf{\bibinfo{volume}{107}},
  \bibinfo{pages}{3905} (\bibinfo{year}{1997}).

\bibitem[{\citenamefont{Hsiao and Zerner}(1999)}]{HZ99}
\bibinfo{author}{\bibfnamefont{Y.-W.} \bibnamefont{Hsiao}} \bibnamefont{and}
  \bibinfo{author}{\bibfnamefont{M.~C.} \bibnamefont{Zerner}},
  \bibinfo{journal}{Int. J. Quantum Chem.} \textbf{\bibinfo{volume}{75}},
  \bibinfo{pages}{577} (\bibinfo{year}{1999}).

\bibitem[{\citenamefont{Bratt et~al.}(2000)\citenamefont{Bratt, Poluektov,
  Thurnauer, Krzystek, Brunel, Schrier, Hsiao, Zerner, and
  Angerhofer}}]{BPM+00}
\bibinfo{author}{\bibfnamefont{P.~J.} \bibnamefont{Bratt}},
  \bibinfo{author}{\bibfnamefont{O.~G.} \bibnamefont{Poluektov}},
  \bibinfo{author}{\bibfnamefont{M.~C.} \bibnamefont{Thurnauer}},
  \bibinfo{author}{\bibfnamefont{J.}~\bibnamefont{Krzystek}},
  \bibinfo{author}{\bibfnamefont{L.-C.} \bibnamefont{Brunel}},
  \bibinfo{author}{\bibfnamefont{J.}~\bibnamefont{Schrier}},
  \bibinfo{author}{\bibfnamefont{Y.-W.} \bibnamefont{Hsiao}},
  \bibinfo{author}{\bibfnamefont{M.}~\bibnamefont{Zerner}}, \bibnamefont{and}
  \bibinfo{author}{\bibfnamefont{A.}~\bibnamefont{Angerhofer}},
  \bibinfo{journal}{J. Phys. Chem. B} \textbf{\bibinfo{volume}{104}},
  \bibinfo{pages}{6973} (\bibinfo{year}{2000}).

\bibitem[{\citenamefont{Rodina et~al.}(2002)\citenamefont{Rodina, Efros, Rosen,
  and Meyer}}]{RER+02}
\bibinfo{author}{\bibfnamefont{A.~V.} \bibnamefont{Rodina}},
  \bibinfo{author}{\bibfnamefont{A.~L.} \bibnamefont{Efros}},
  \bibinfo{author}{\bibfnamefont{M.}~\bibnamefont{Rosen}}, \bibnamefont{and}
  \bibinfo{author}{\bibfnamefont{B.~K.} \bibnamefont{Meyer}},
  \bibinfo{journal}{Mat. Sci. Eng. C} \textbf{\bibinfo{volume}{19}},
  \bibinfo{pages}{435} (\bibinfo{year}{2002}).

\bibitem[{\citenamefont{Carter et~al.}(1997)\citenamefont{Carter, Bouldin,
  Kemner, Bell, Woicik, and Majetich}}]{CBK+97}
\bibinfo{author}{\bibfnamefont{A.~C.}~\bibnamefont{Carter}},
  \bibinfo{author}{\bibfnamefont{C.~E.} \bibnamefont{Bouldin}},
  \bibinfo{author}{\bibfnamefont{K.~M.} \bibnamefont{Kemmer}},
  \bibinfo{author}{\bibfnamefont{M.~I.} \bibnamefont{Bell}},
  \bibinfo{author}{\bibfnamefont{J.~C.} \bibnamefont{Woicik}},
  \bibnamefont{and} \bibinfo{author}{\bibfnamefont{S.~A.}
  \bibnamefont{Majetich}}, \bibinfo{journal}{Phys. Rev. B}
  \textbf{\bibinfo{volume}{55}}, \bibinfo{pages}{13822} (\bibinfo{year}{1997}).

\bibitem[{\citenamefont{Hu et~al.}(2002)\citenamefont{Hu, Wang, Li, Yang, and
  Alivisatos}}]{HWL+02}
\bibinfo{author}{\bibfnamefont{J.}~\bibnamefont{Hu}},
  \bibinfo{author}{\bibfnamefont{L.-W.} \bibnamefont{Wang}},
  \bibinfo{author}{\bibfnamefont{L.-S.} \bibnamefont{Li}},
  \bibinfo{author}{\bibfnamefont{W.}~\bibnamefont{Yang}}, \bibnamefont{and}
  \bibinfo{author}{\bibfnamefont{A.~P.} \bibnamefont{Alivisatos}},
  \bibinfo{journal}{J. Phys. Chem. B} \textbf{\bibinfo{volume}{106}},
  \bibinfo{pages}{2447} (\bibinfo{year}{2002}).

\bibitem[{\citenamefont{Hu et~al.}(2001)\citenamefont{Hu, Li, Yang, Manna,
  Wang, and Alivisatos}}]{HLY+01}
\bibinfo{author}{\bibfnamefont{J.}~\bibnamefont{Hu}},
  \bibinfo{author}{\bibfnamefont{L.~S.} \bibnamefont{Li}},
  \bibinfo{author}{\bibfnamefont{W.~D.} \bibnamefont{Yang}},
  \bibinfo{author}{\bibfnamefont{L.}~\bibnamefont{Manna}},
  \bibinfo{author}{\bibfnamefont{L.~W.} \bibnamefont{Wang}}, \bibnamefont{and}
  \bibinfo{author}{\bibfnamefont{A.~P.} \bibnamefont{Alivisatos}},
  \bibinfo{journal}{Science} \textbf{\bibinfo{volume}{292}},
  \bibinfo{pages}{2060} (\bibinfo{year}{2001}).

\bibitem[{\citenamefont{Kadavanich}(1997)}]{Kad97}
\bibinfo{author}{\bibfnamefont{A.~V.} \bibnamefont{Kadavanich}}, Ph.D. thesis,
  \bibinfo{school}{University of California, Berkeley} (\bibinfo{year}{1997}).

\bibitem[{\citenamefont{Vogl et~al.}(1983)\citenamefont{Vogl, Hjalmarson, and
  Dow}}]{VHD83}
\bibinfo{author}{\bibfnamefont{P.}~\bibnamefont{Vogl}},
  \bibinfo{author}{\bibfnamefont{H.~P.} \bibnamefont{Hjalmarson}},
  \bibnamefont{and} \bibinfo{author}{\bibfnamefont{J.~D.} \bibnamefont{Dow}},
  \bibinfo{journal}{J. Phys. Chem. Solids} \textbf{\bibinfo{volume}{44}},
  \bibinfo{pages}{365} (\bibinfo{year}{1983}).

\bibitem[{\citenamefont{Troparevsky et~al.}(2001)\citenamefont{Troparevsky,
  Kronik, and Chelikowsky}}]{TKC01}
\bibinfo{author}{\bibfnamefont{M.~C.} \bibnamefont{Troparevsky}},
  \bibinfo{author}{\bibfnamefont{L.}~\bibnamefont{Kronik}}, \bibnamefont{and}
  \bibinfo{author}{\bibfnamefont{J.~R.} \bibnamefont{Chelikowsky}},
  \bibinfo{journal}{Phys. Rev. B} \textbf{\bibinfo{volume}{65}},
  \bibinfo{pages}{033311} (\bibinfo{year}{2001}).

\bibitem[{\citenamefont{Patchkovskii and Ziegler}(2001)}]{PZ01}
\bibinfo{author}{\bibfnamefont{S.}~\bibnamefont{Patchkovskii}}
  \bibnamefont{and} \bibinfo{author}{\bibfnamefont{T.}~\bibnamefont{Ziegler}},
  \bibinfo{journal}{J. Phys. Chem. A} \textbf{\bibinfo{volume}{105}},
  \bibinfo{pages}{5490} (\bibinfo{year}{2001}).

\end{thebibliography}
%\bibliographystyle{apsrev}

\vfill\eject

\begin{table}
\caption{Number of $g$ factor components observed in TRFR experiments
\cite{GAP+99, GAE+02} as a function of nanocrystal effective radius
and aspect ratio.  For the 22 \AA~ and 25 \AA~ dots we use the
sixth-order polynomial fit to aspect ratio described in
Ref. \onlinecite{LPW98}; for the 40 \AA~ and 50 \AA~ dots we use the
linear fit described in Ref. \onlinecite{Kad97}; for the 80 \AA~ dot
we use the aspect ratio given in Ref. \onlinecite{GAE+02}.}
\label{gcomponent_table}
\begin{tabular}{|c|c|c|}\hline
Effective Radius (\AA) & Aspect Ratio & $g$ components \\
\hline
22 & 1.17 & 1 \\
\hline
25 & 1.20 & 2 \\
\hline
40 & 1.23 & 2 \\
\hline
57 & 1.34 & 4 \\
\hline
80 & 2 & 2 \\
\hline
\end{tabular}
\end{table}

\newpage

\begin{figure}
\includegraphics[width=4in]{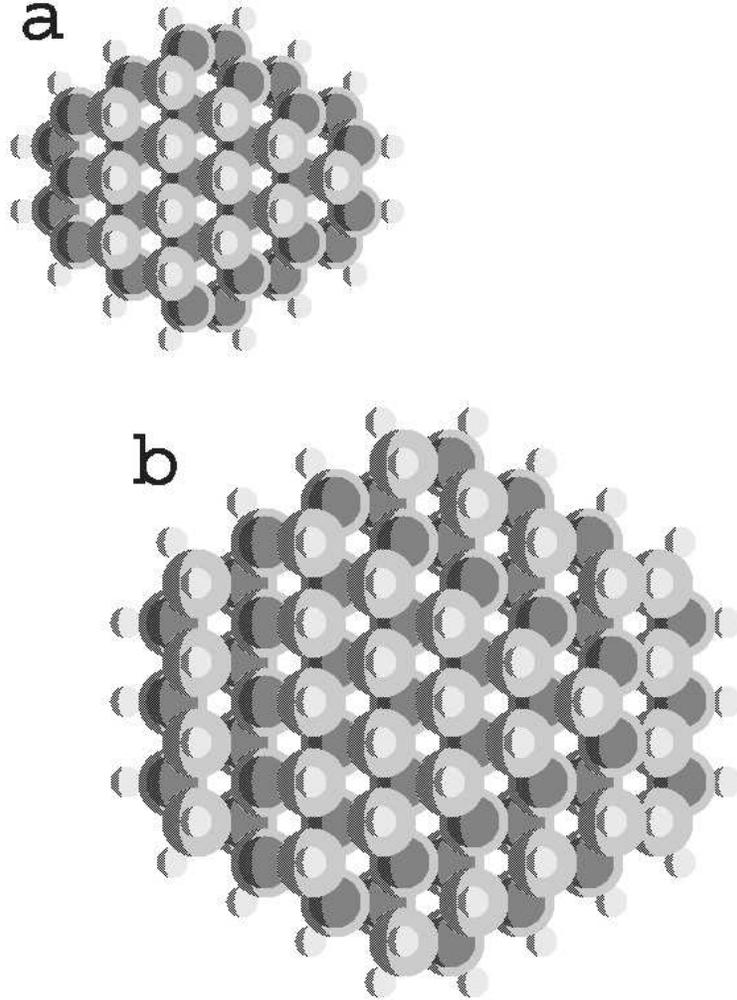}
\caption{Cross section of nanocrystal rods.  Small circles are ligand atoms.
Large light circles are Cd atoms, and medium dark circles are Se
atoms. a) Cross section of \smallrod~ (small) rod.  
b) Cross section of \bigrod~ (large) rod.}
\end{figure}

\begin{figure}
\label{ndoped_dang_fig}
\includegraphics[width=6in]{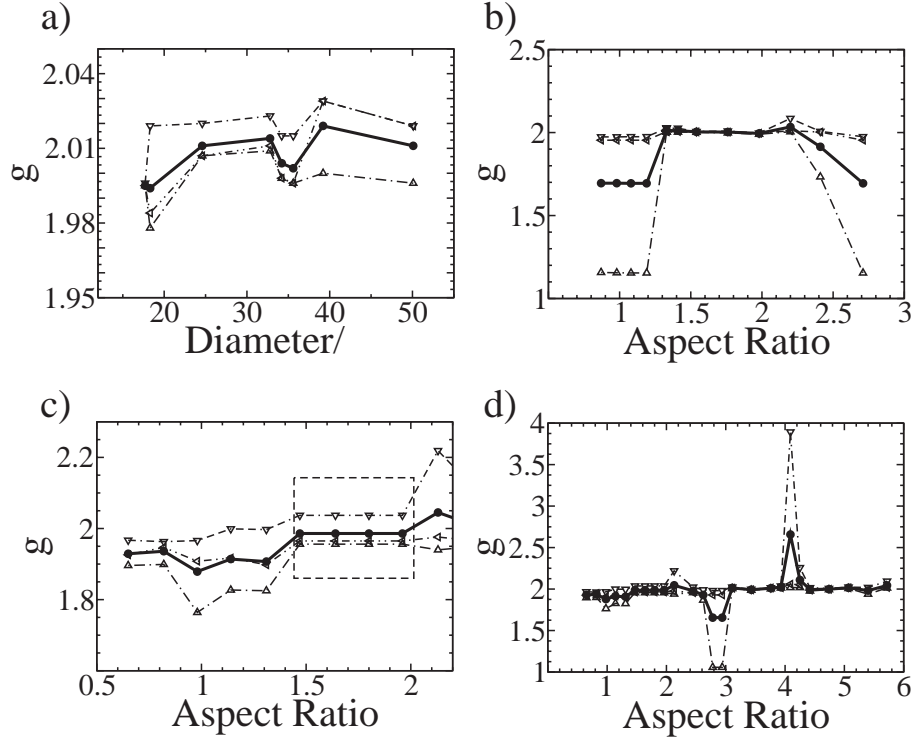}
\caption{Conduction electron $g$ factors for the $n$-doped CdSe nanocrystals.
In all figures, dashed lines with open symbols represent the three anisotropic
components of the $g$ tensor, and the solid line with filled circles represents
the geometric mean (``isotropic'' $g$ factor) of the components.  
a) $n$-doped dot $g$ factors as a function of dot diameter in Angstroms; the
number of atoms varies from 96 to 1783.
b) \bigrod~ cross-section $n$-doped rod $g$ factors as a
function of aspect ratio; the number of atoms varies from 705 to 2252.  
Note the clear discontinuity in the
$g$ factor at aspect ratio $\sim 1.3$, and the quasi-spherical region
extending between aspect ratio 1.3 and 2.
c) \smallrod~ cross-section $n$-doped rod $g$ factors as a
function of aspect ratio; the number of atoms varies from 197 to 651.
Note the jump in the $g$ factor
$\Delta g_{iso} \sim 0.1$ at aspect ratio $\sim 1.3$ and the 
quasi-spherical region (shown in dashed box).
d) \smallrod~ cross section $n$-doped rod $g$ factors over
a longer range of aspect ratios; the number of atoms varies from 197 to
1773.  }
\end{figure}

\begin{figure}
\label{exciton_dang_fig}
\includegraphics[width=6in]{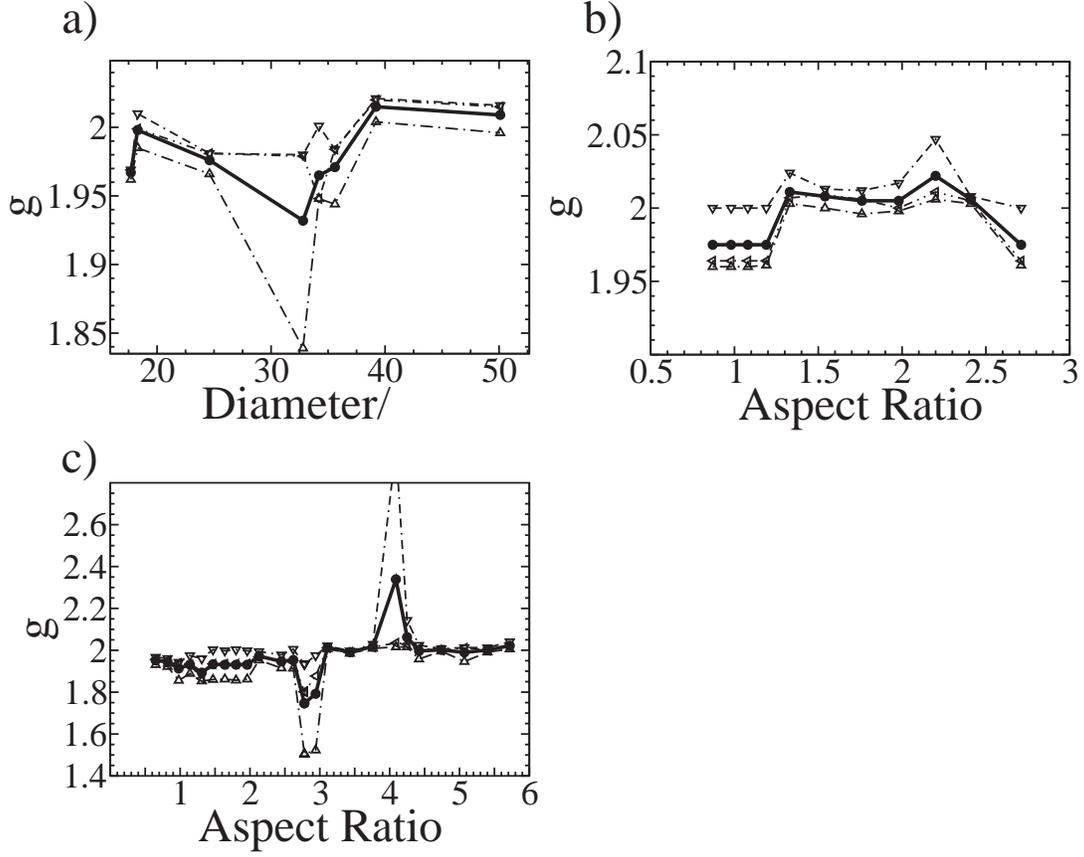}
\caption{$g$ factors in CdSe nanocrystals, calculated for electronic
  configurations having two unpaired electrons with parallel spin in
  valence and conduction band edge states.  
In all figures, dashed lines with open symbols represent the three anisotropic
components of the $g$ tensor, and the solid line with filled circles represents
the geometric mean (``isotropic'' $g$ factor) of the components.  
See caption of Figure 2 for corresponding number of atoms.
Comparing to the previous
figure, the qualitative behavior is similar in all cases. a) Dots.  
b) \bigrod~ cross-section rods.  Note that while the
qualitative behavior is similar, the magnitude of the $g$ factor
discontinuity is reduced by an order of magnitude.
c) \smallrod~ cross-section rods over a larger range of aspect ratios. }
\end{figure}

\begin{figure}
\label{ndoped_nodang_fig}
\includegraphics[width=6in]{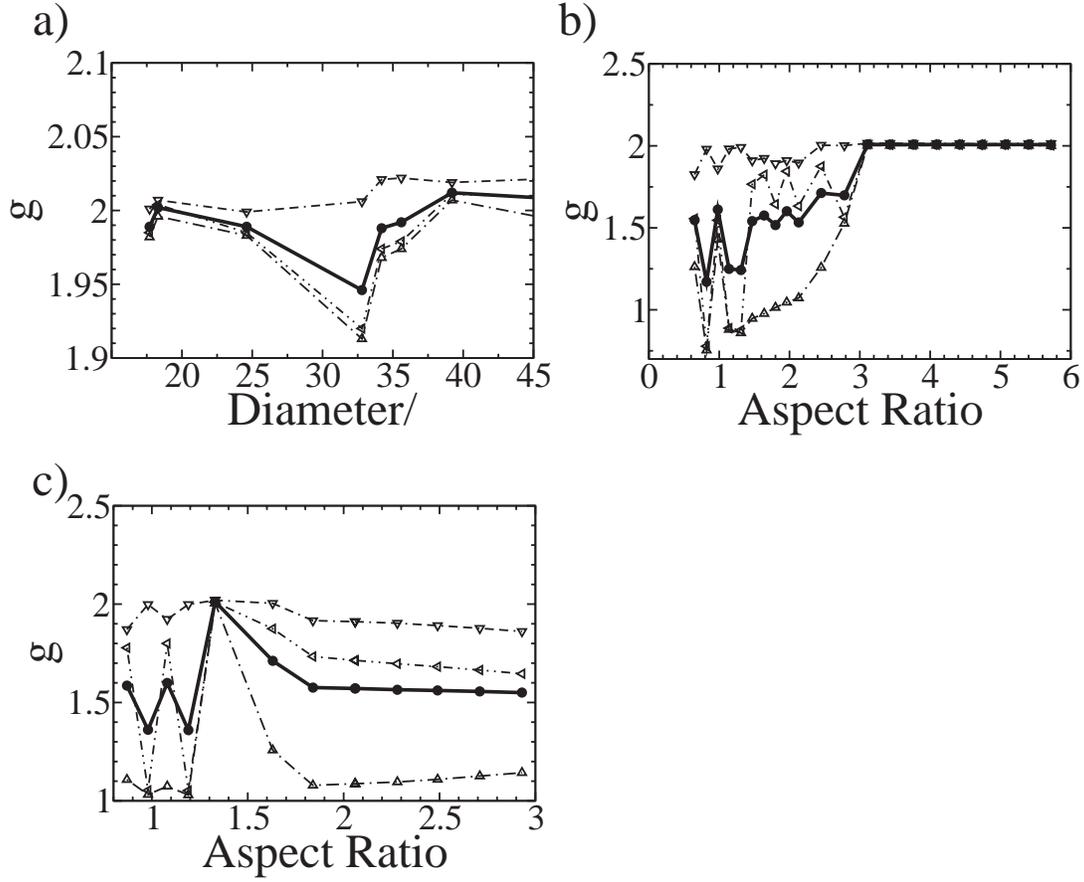}
\caption{Conduction electron $g$ factors for $n$-doped CdSe
  nanocrystals with Se dangling
bonds truncated.
In all figures, dashed lines with open symbols represent the three anisotropic
components of the $g$ tensor, and the solid line with filled circles represents
the geometric mean (``isotropic'' $g$ factor) of the components.
See caption of Figure 2 for corresponding number of atoms.
   a) Dots. The decrease in the $g$ factor magnitudes
for the 35 \AA~ diameter dots is increased as compared to the dangling
bond calculations, but the overall qualitative behavior is unchanged.
  b) \bigrod~ cross-section
rods. Note the abrupt change at aspect ratio $\sim 1.3$, but the lack
of an isotropic behavior between aspect ratios $1.3-2$ as seen in the
dangling bond calculation in Figure 2b.  
c) \smallrod~ cross-section rods.  These small rods
show qualitatively different behavior, with the $g$ factor becoming
isotropic at significantly larger aspect ratios, greater than $\sim 3$.}
\end{figure}

\begin{figure}
\label{exciton_nodang_fig}
\includegraphics[width=6in]{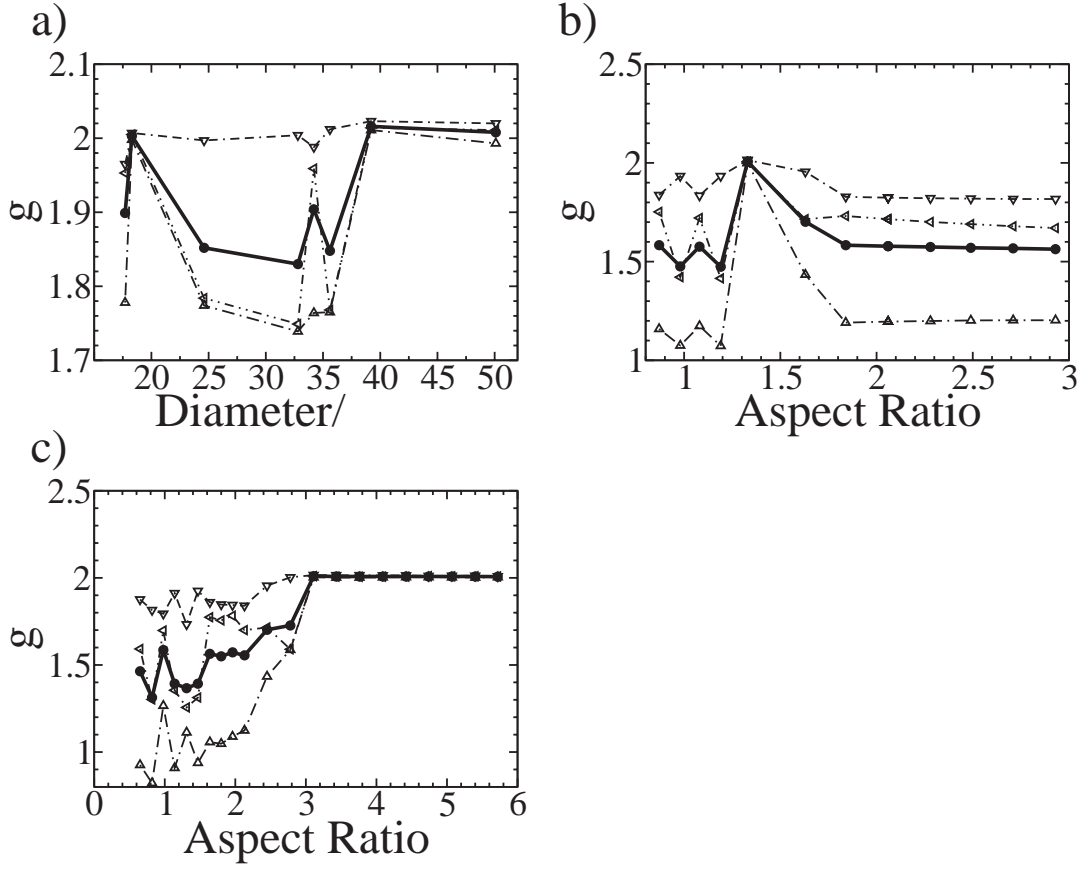}
\caption{$g$ factors for neutral CdSe nanocrystals with Se dangling
bonds truncated.  All calculations are made for electronic
configurations having two unpaired electrons with parallel spin in
valence and conduction band edge states.  
In all figures, dashed lines with open symbols represent the three anisotropic
components of the $g$ tensor, and the solid line with filled circles represents
the geometric mean (``isotropic'' $g$ factor) of the components.
See caption of Figure 2 for corresponding number of atoms.
a) Dots.  b) \bigrod~ cross-section
rods. c) \smallrod~ cross-section rods.  Note in all cases
the qualitative similarity to the conduction band electron
$g$ factors shown in Figure 4.}
\end{figure}

\begin{figure}
\label{rod2_character}
\includegraphics[width=6in]{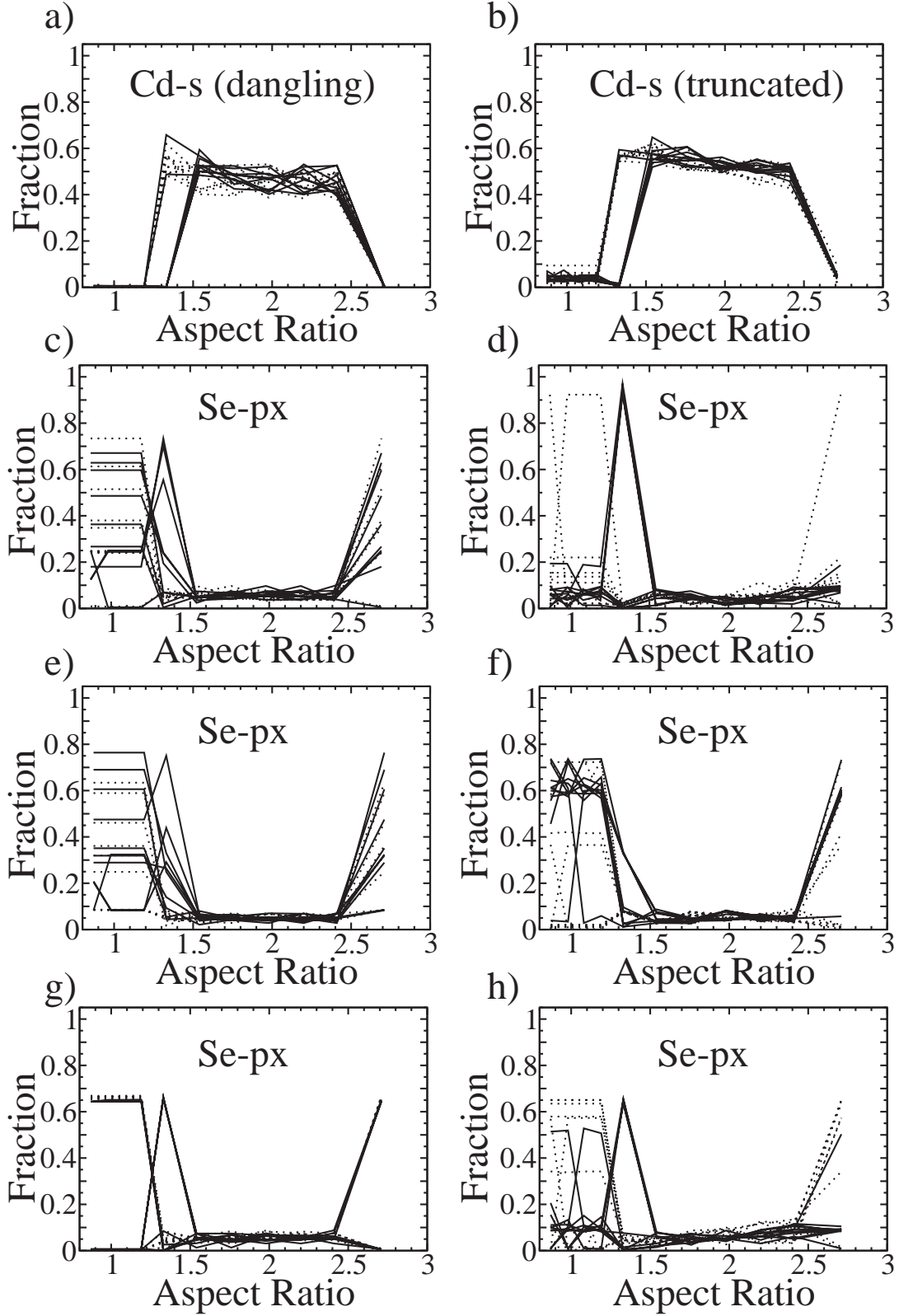}
\caption{Fractional atomic-orbital contents for the \bigrod~
cross-section CdSe nanocrystal rods, as a function of aspect ratio.
The number of atoms varies from 705 to 2252.  Dotted lines depict the
content of the conduction band edge and the 9 levels above; solid
lines depict the content of the valence band edge and the 9 levels
below.  Orbital types where the maximum fractional content was less
than 0.15 are omitted.  Left column panels are with the inclusion of Se
dangling bonds, right column panels  truncate Se dangling bonds. 
\iffalse
a) Cd-$s$
content (dangling).  b) Cd-$s$ content (truncated). c) Se-$p_{x}$
content (dangling).  d) Se-$p_{x}$ content (truncated).  e) Se-$p_{y}$
content (dangling).  f) Se-$p_{y}$ content (truncated).  g) Se-$p_{z}$
content (dangling).  h) Se-$p_{z}$ content (truncated).
\fi
}
\end{figure}

\end{document}